\documentclass[12pt,preprint]{aastex}
\usepackage{natbib}

\begin{document}

\title{ Magneto-convection in a sunspot umbra}
\author{M. Sch\"ussler and A. V\"ogler\altaffilmark{1}}
\affil{Max Planck Institut f\"ur Sonnensystemforschung, 
       Max-Planck-Str.~2,
       37191 Katlenburg-Lindau, Germany}

\altaffiltext{1}{Current address: High Altitude Observatory,
NCAR, P.O. Box 3000, Boulder, Colorado 80307, USA}

\email{schuessler@mps.mpg.de, voegler@mps.mpg.de}

\begin{abstract}
Results from a realistic simulation of 3D radiative magneto-convection
in a strong background magnetic field corresponding to the conditions in
sunspot umbrae are shown. The convective energy transport is dominated by
narrow upflow plumes with adjacent downflows, which become almost
field-free near the surface layers. The strong external magnetic field
forces the plumes to assume a cusp-like shape in their top parts, where
the upflowing plasma loses its buoyancy. The resulting bright features
in intensity images correspond well (in terms of brightness, size, and
lifetime) to the observed umbral dots in the central parts of sunspot
umbrae. Most of the simulated umbral dots have a horizontally elongated
form with a central dark lane. Above the cusp, most plumes show narrow
upflow jets, which are driven by the pressure of the piled-up plasma
below. The large velocities and low field strengths in the plumes are
effectively screened from spectroscopic observation because the surfaces
of equal optical depth are locally elevated, so that spectral lines are
largely formed above the cusp. Our simulations demonstrate that nearly
field-free upflow plumes and umbral dots are a natural result of
convection in a strong, initially monolithic magnetic field.
\end{abstract}


\section{Introduction}

Sunspot umbrae appear dark because the convective flows are suppressed
by the strong magnetic field \citep{Biermann:1941, Cowling:1953},
reducing the emitted energy (per unit time and area) in the central
parts of a large umbra to about 10--20\% of the average value outside
sunspots. However, even this strongly diminished energy flux cannot be
carried by radiation alone below the umbral photosphere
\citep{Schlueter:Temesvary:1958}, so that some form of reduced
convective energy transport is required to construct a consistent
sunspot model \citep{Deinzer:1965}. The observation of umbral dots,
relatively bright features of sub-arcsecond size embedded in the dark
umbral background, has often been taken as a signature of such
convective energy transport, either in the form of overstable
oscillations (`elevator convection') in thin vertical columns
\citep[e.g.,][]{Savage:1969} or as intrusions from below of non-magnetic
plasma into a shallow cluster-type sunspot \citep{Parker:1979a}. For
monolithic sunspot models, linear stability analysis indicates that
oscillatory convection in slender columns is the preferred mode in the
first few Mm depth below the umbral photosphere, where the quantity
$\zeta=\eta/\kappa$, the ratio of the magnetic diffusivity, $\eta$, to
the thermal (radiative) diffusivity, $\kappa$, has values smaller than
unity \citep{Meyer:etal:1974}. Systematic studies based on numerical
simulations under idealized conditions extend this result into the
nonlinear regime \citep[e.g.,][]{Weiss:etal:1990,Weiss:etal:1996,
Weiss:etal:2002}. Here we report on the first realistic simulations of
convection in sunspot umbrae with full radiative transfer and partial
ionization effects. These simulations can be compared directly with
observations.

\section{Simulation setup}

We have carried out ab-initio numerical simulations of 3D radiative
magneto-con\-vect\-ion in the near-surface layers of a sunspot umbra
using the {\sl MURaM} code\footnote{
http://www.mps.mpg.de/projects/solar-mhd/muram\_site}
\citep{Voegler:etal:2003,Voegler:2003, Voegler:etal:2005}. The
computational box extends $5760\,$km$\times 5760\,$km in the horizontal
directions and $1600\,$km in depth, roughly covering the range from
$1200\,$km below to $400\,$km above the average level of Rosseland
optical depth unity, $\tau_{\mathrm R}=1$. The cell size of the
computational mesh is $20\,$km in both horizontal directions and
$10\,$km in the vertical. We assume a constant magnetic diffusivity of
2.8$\cdot10^6$m$^2$s$^{-1}$ and hyperdiffusivities for the other
diffusive processes \citep[for details, see][]{Voegler:etal:2005}.
Since here we are mainly interested in studying the convective processes
around and below $\tau_{\mathrm R}=1$ without a detailed comparison with
spectro-polarimetric observations, we have included only the lower parts
of the umbral atmosphere and also restricted ourselves to grey radiative
transfer. The vertical magnetic flux through the box is fixed,
corresponding to a horizontally averaged vertical magnetic field of
$2500\,$G. The thermal energy density of the inflowing matter at the
(open) bottom boundary has been fixed at a value of
$3.5\cdot10^{12}\,$erg$\cdot$cm$^{-3}$, leading to an average radiative
energy output (per unit area and time) through the upper boundary around
17--18\% of its value outside sunspots.

The first phase of the thermal relaxation of the model was carried out
in 2D for larger computational efficiency. After about 30 hours of
computed solar time, we continued to evolve the model in 3D for about 3
hours until a statistically stationary state (i.e., no trends in
energy flux and in horizontally averaged profiles of temperature and
other quantities) had developed. The simulation was then continued for
another 2.3 hours of solar time.

\section{Results}

\begin{figure*}[ht!]
\includegraphics[scale=0.57]{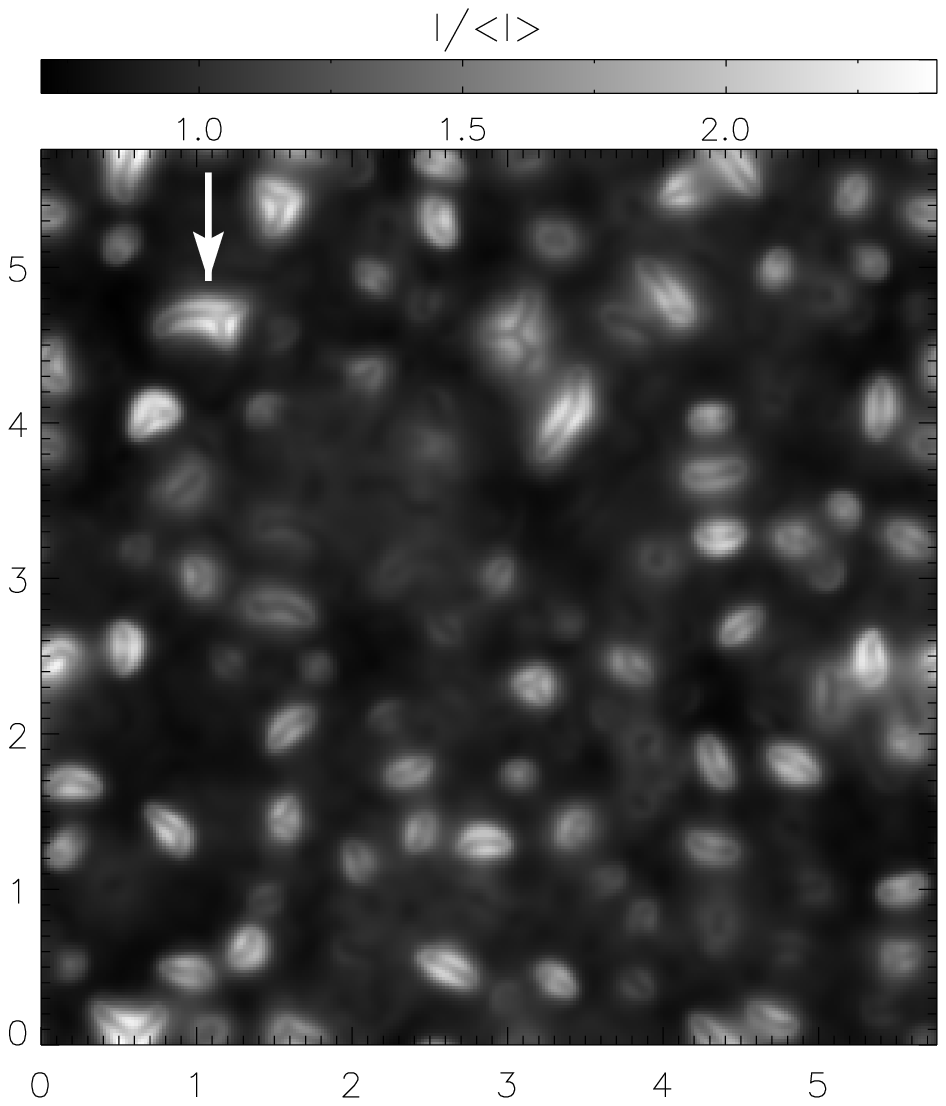}\hspace{-3mm}
\includegraphics[scale=0.57]{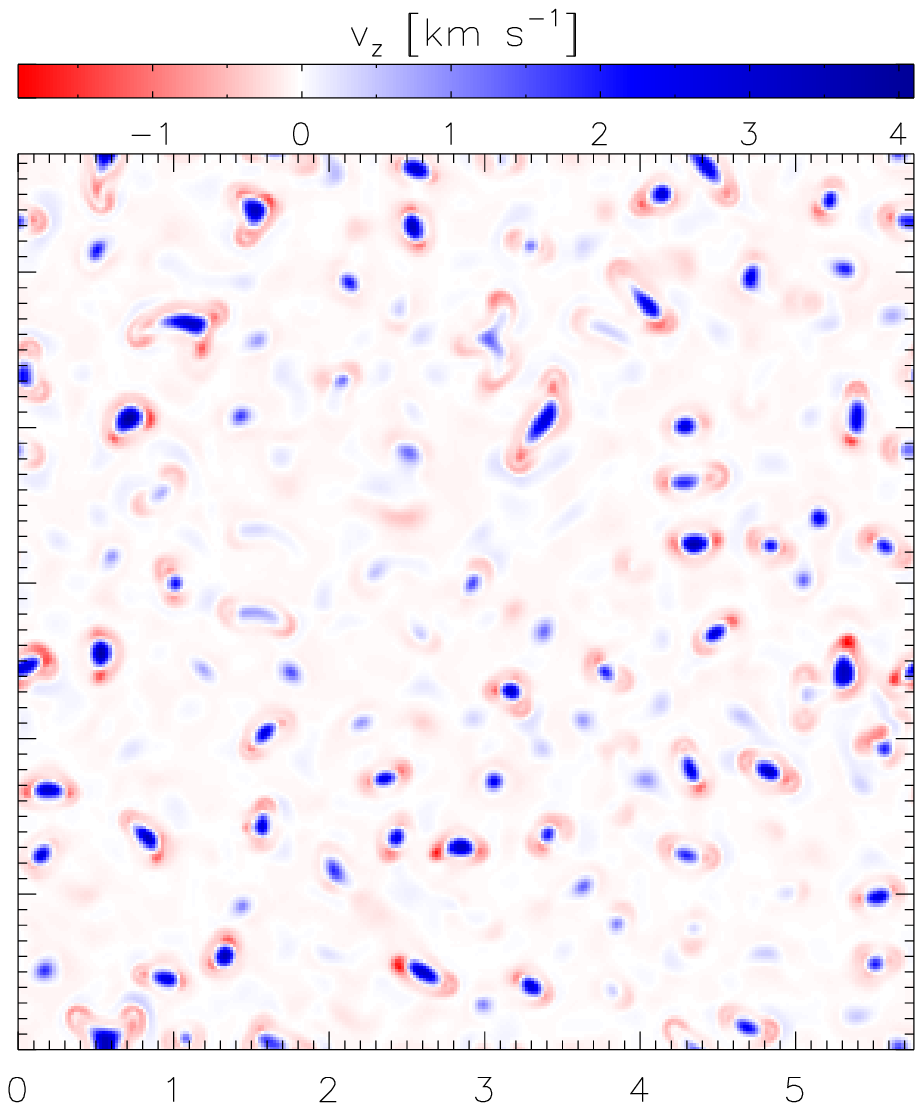}\hspace{-3.5mm}
\includegraphics[scale=0.57]{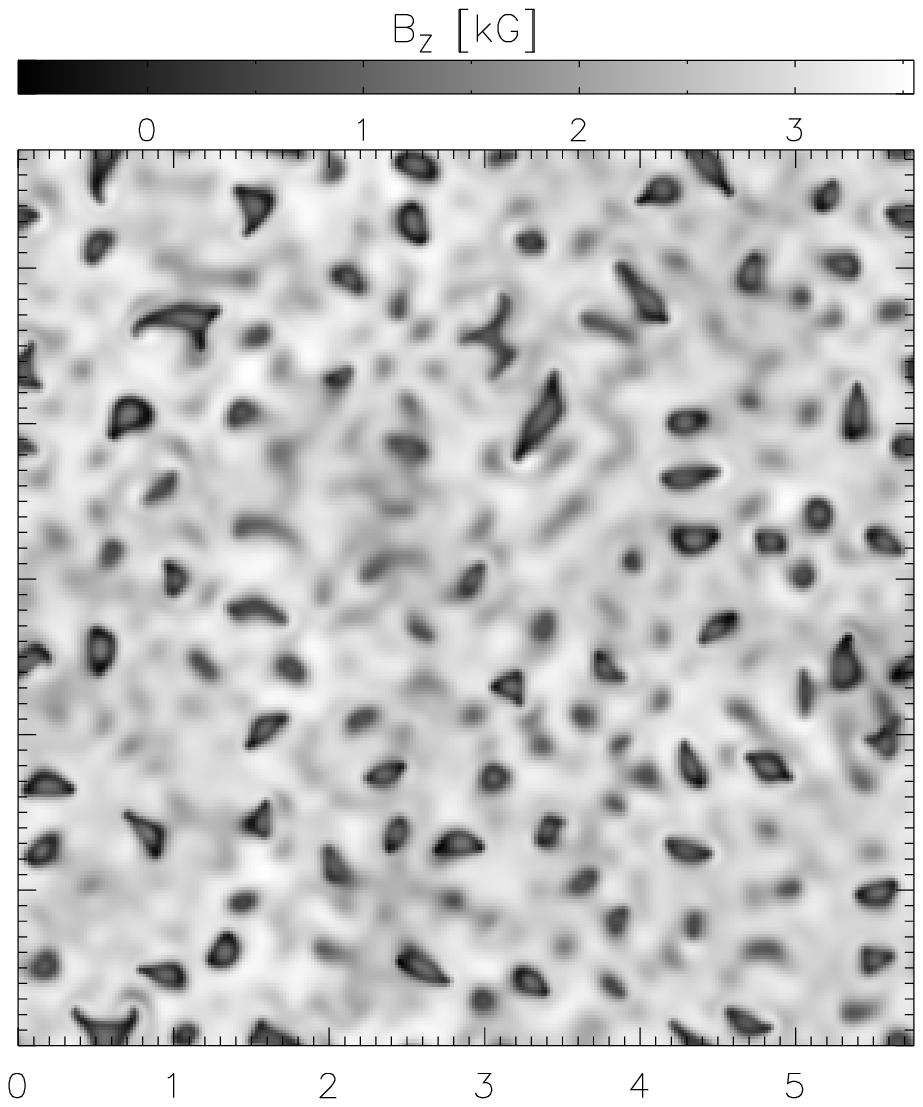}
\caption{Vertically emerging grey intensity, normalized by its
 horizontal average (left) and cuts of the vertical velocity (middle)
 and vertical magnetic field (right) components at a height of
 $1200\,$km above the bottom of the simulation box ($400\,$km below the
 top), approximately corresponding to the average level of Rosseland
 optical depth unity ($\tau_{\mathrm R}=1$), for a snapshot from the
 simulation run. The length unit is Mm. The bright features in the
 intensity image can be identified with umbral dots. They are caused by
 strong upflows in regions of significantly reduced magnetic field. The
 arrow indicates the umbral dot studied in more detail below.}
\label{fig:slices}
\end{figure*}

The system develops a mode of convective energy transport which is
dominated by non-stationary narrow plumes of rising hot plasma. The
strong expansion of the rising matter with height leads to a drastic
decrease of the magnetic field strength in the upper layers of the
plumes.  For a snapshot from the simulation run, Fig.~\ref{fig:slices}
shows the brightness (vertically emerging grey intensity) together with
the vertical velocity and the vertical magnetic field, the latter two at
a height of $z=1200\,$km, which roughly corresponds to the average level
of $\tau_{\mathrm R}=1$. The bright features in the intensity image have
a typical size of a 200--300 kilometers, a lifetime of the order of 30
minutes, and a broad distribution of brightness values reaching up to
2.4 times the average brightness in the simulated umbra. These values
are consistent with the results of high-resolution photometry of umbral
dots in the darker central parts of sunspot umbrae
\citep{Sobotka:Hanslmeier:2005}.  Most of the simulated umbral dots have
an elongated shape with a central dark lane and downflows concentrated
at the end points. Larger dots sometimes show a threefold dark lane with
three downflow patches. These shapes possibly result from the fluting
instability in the upper parts of the low-field-strength upflow channel
underlying the umbral dot: the elongated features correspond to
azimuthal wavenumber $m=2$ while the threefold structure results from
the mode $m=3$.

\begin{figure}[htb]
\includegraphics[scale=0.9]{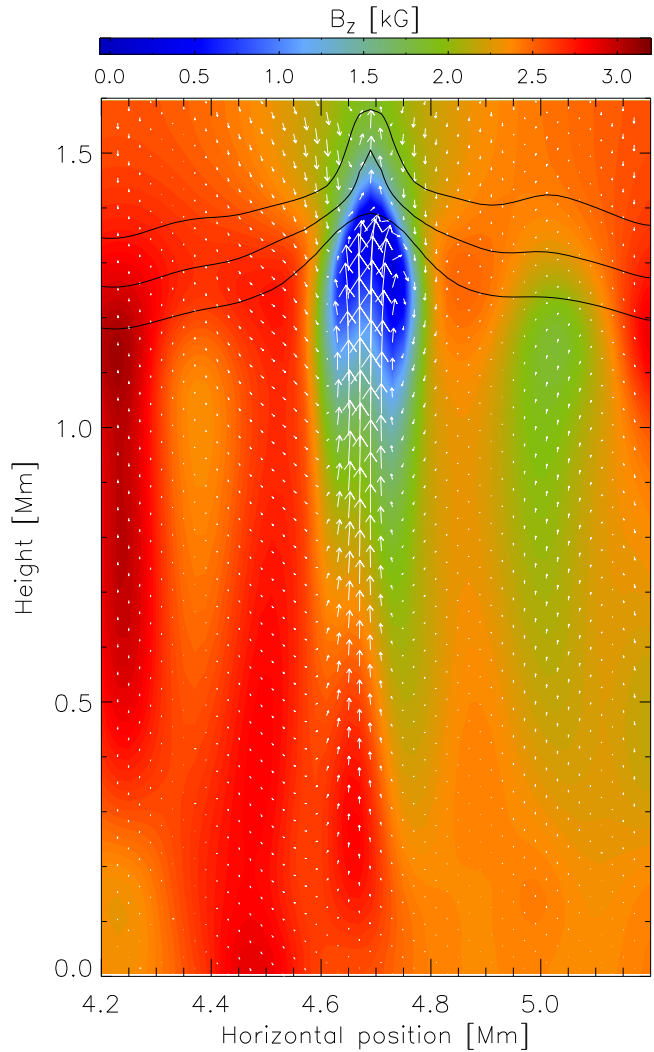}
\caption{Vertical cut through the simulation box in the direction
         indicated by the arrow in Fig.~\ref{fig:slices}, nearly
         perpendicular to the dark lane of the umbral dot. Colours
         indicate magnetic field strength, the arrows represent
         (projected) velocity vectors. The longest arrow corresponds to
         a velocity of $2.7\,$km$\cdot$s$^{-1}$.  The lines indicate the
         levels of constant optical depth $\tau_{\mathrm R}=1.$, 0.1,
         and 0.01, respectively (from bottom to top). The upper part of
         the plume has developed a cusp-like shape with a strongly
         decelerated uflow and a weak magnetic field. The flow turns
         horizontal mainly in the direction perpendicular to the plane
         of the cut.}
\label{fig:cut_y}
\end{figure}

\begin{figure}[htb]
\plotone{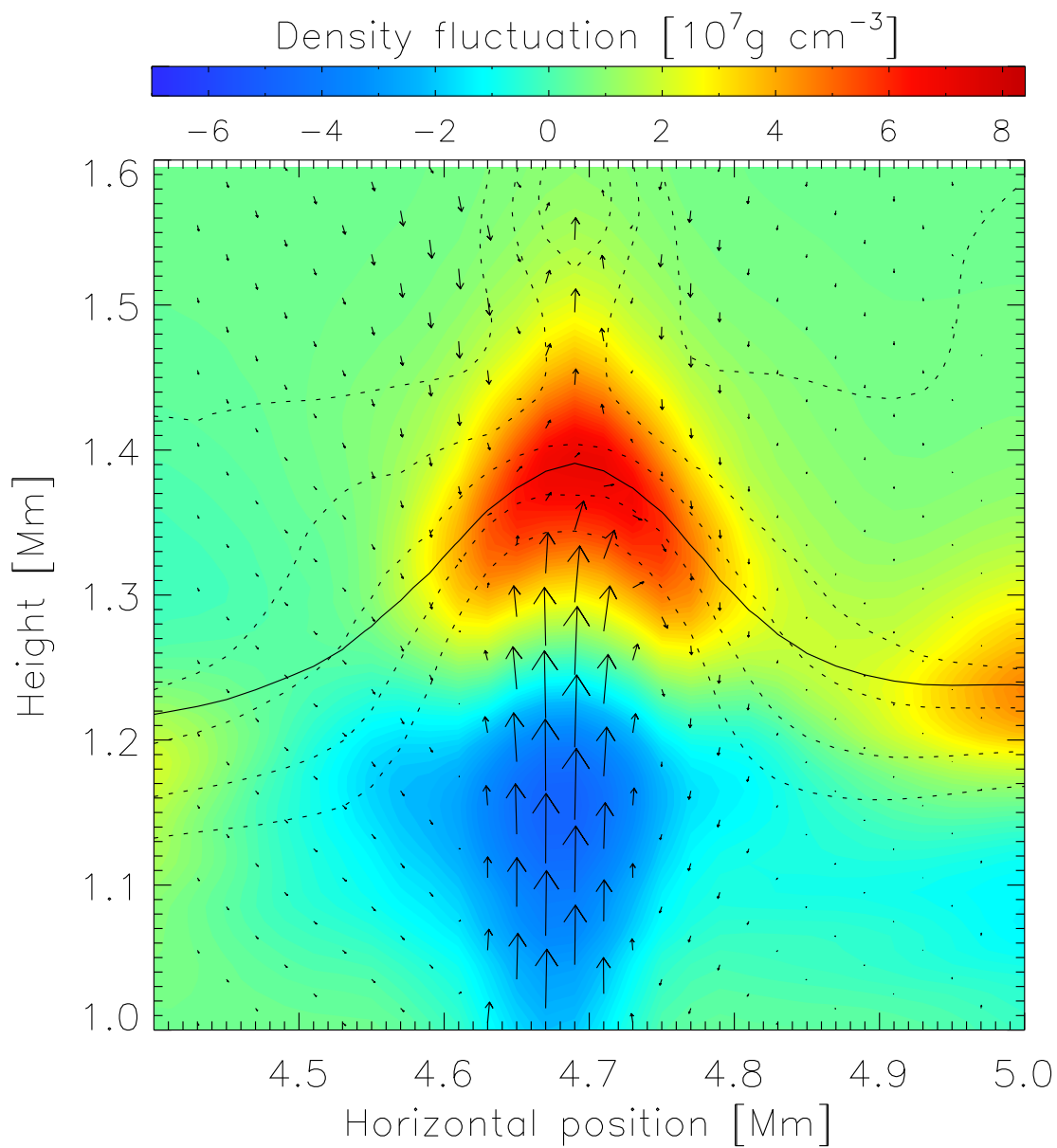}
\caption{Upper central part of the cut shown in Fig.~\ref{fig:cut_y}.
Colours represent the density fluctuation with respect to the horizontal
mean at the same height. The full line indicates the level of
$\tau_{\mathrm R}=1$ while the dashed lines are isotherms. The piling up
of matter raises the surfaces of constant optical depth into
low-temperature regions near the cusp, leading to the appearance of a
dark lane with a maximum contrast of about 15\% in the intensity
image. The longest velocity arrow corresponds to a speed of
$2.7\,$km$\cdot$s$^{-1}$.}
\label{fig:blowup}
\end{figure}

\begin{figure}[htb]
\epsscale{1.00}
\plotone{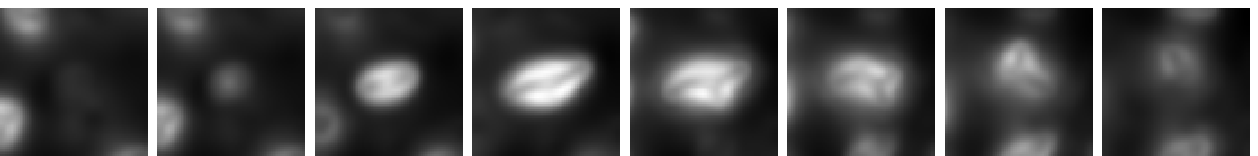}\\
\plotone{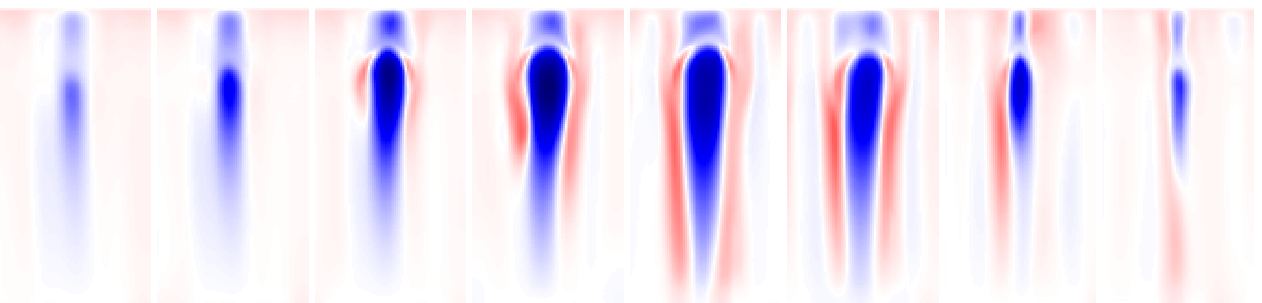}\\
\plotone{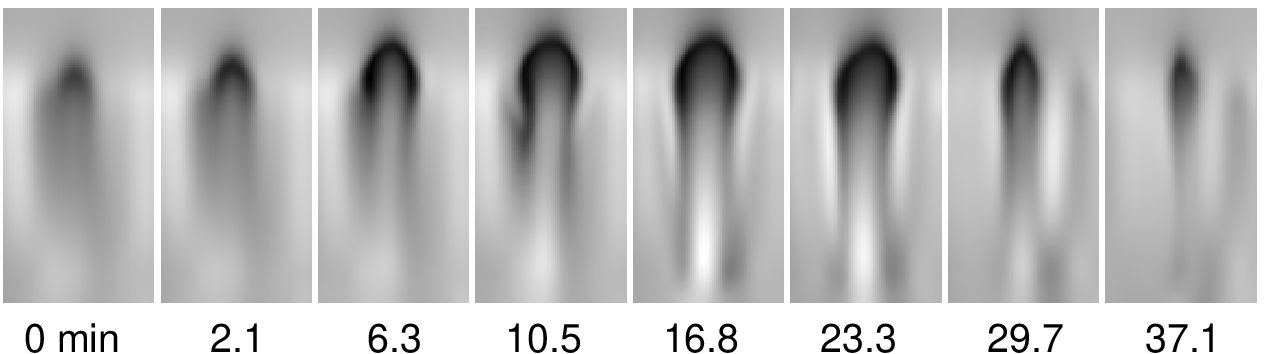}
\caption{Time evolution of an upflow plume developing into an umbral
  dot. The panels show, from left to right, the development of the
  brightness (top row) and of vertical cuts roughly along the dark lane
  in the brightness images. Shown are the vertical velocity (middle row,
  blue: upflow, red: downflow) and the vertical magnetic field (bottom
  row), the cuts covering the full depth of the simulation box and
  $820\,$km in horizontal direction. The colour and greyscale schemes
  are the same as those used in Fig.~\ref{fig:slices}, albeit with
  different ranges: $I/\langle I\rangle = 0.7\dots2.25$, $v_z
  =-1.2\dots4.1\,$km$\cdot$s$^{-1}$, $B_z = -0.3\dots3.8\,$kG. The
  snapshots are taken at the times (in minutes, relative to the first
  snapshot) indicated below the bottom panel. The umbral dot is clearly
  detectable as a bright structure for at least 25 minutes.  }
\label{fig:timeseries}
\end{figure}

We have studied in some detail the structure of the umbral dot indicated
by the arrow in the left panel of Fig.~\ref{fig:slices}. A cut through
the computational box in the direction of the arrow, roughly
perpendicular to the dark lane, is shown in Fig.~\ref{fig:cut_y}. The
cut covers the upflow plume underlying the umbral dot and its immediate
environment.
The magnetic field is strongly reduced down to values of a few hundred
Gauss in the near-surface layers of the plume. The corresponding lateral
expansion of the structure is mainly in the direction perpendicular to
the cut shown here, leading to the elongated shape of the visible umbral
dot.  The upflow is strongly braked around $\tau_{\mathrm R}=1$, where
the plasma rapidly loses its buoyancy, mainly owing to cooling by
radiative losses and the mean stratification becoming subadiabatic. The
plasma piles up and the flow turns horizontal (mainly in the direction
perpendicular to the cut) in the region below the cusp until a
quasi-stationary state with locally enhanced density and pressure is
established. This is illustrated in Fig.~\ref{fig:blowup}, which shows
the density fluctuations with respect to the horizontal average at the
same height in the upper central part of the cut displayed in
Fig.~\ref{fig:cut_y}. As a result of the enhanced density below the
cusp, the level of $\tau_{\mathrm R}=1$ (full line) is so much elevated
in the central part that it cuts through regions of lower temperature
(indicated by the dashed isotherms) than in the more peripheral
parts. This leads to the dark lane in the intensity image; a similar
mechanism has been proposed to explain the dark lanes observed in bright
penumbral filaments \citep{Spruit:Scharmer:2006}. The flow turns
horizontal near the surface of $\tau_{\mathrm R}=1$ and proceeds in the
direction of the dark lane, finally descending in narrow downflow
channels, which are still within the volume that the flow has largely
cleared from magnetic field (cf. Fig.~\ref{fig:slices}, see also
Fig.~\ref{fig:timeseries}).


An interesting feature visible in Figs.~\ref{fig:cut_y} and
\ref{fig:blowup} is a narrow jet-like upflow above the cusp. The
pressure below the cusp builds up sufficiently strongly to drive matter
out along the magnetic field above the cusp with a velocity of about
$1\,$km$\cdot$s$^{-1}$, the cusp acting like a safety valve
\citep{Choudhuri:1986}. Although most of the upflowing matter turns over
and descends in the downflow channels and only a small part escapes
upward, such outflows occur above most upflow plumes (see also
Fig.~\ref{fig:timeseries}) and could possibly affect the chromospheric
dynamics above sunspots. However, since the simulation does not cover
the higher layers of the umbral atmosphere and has a closed top boundary,
we cannot predict the effects on the basis of the present results.

\begin{figure}[htb]
\epsscale{.7}
\plotone{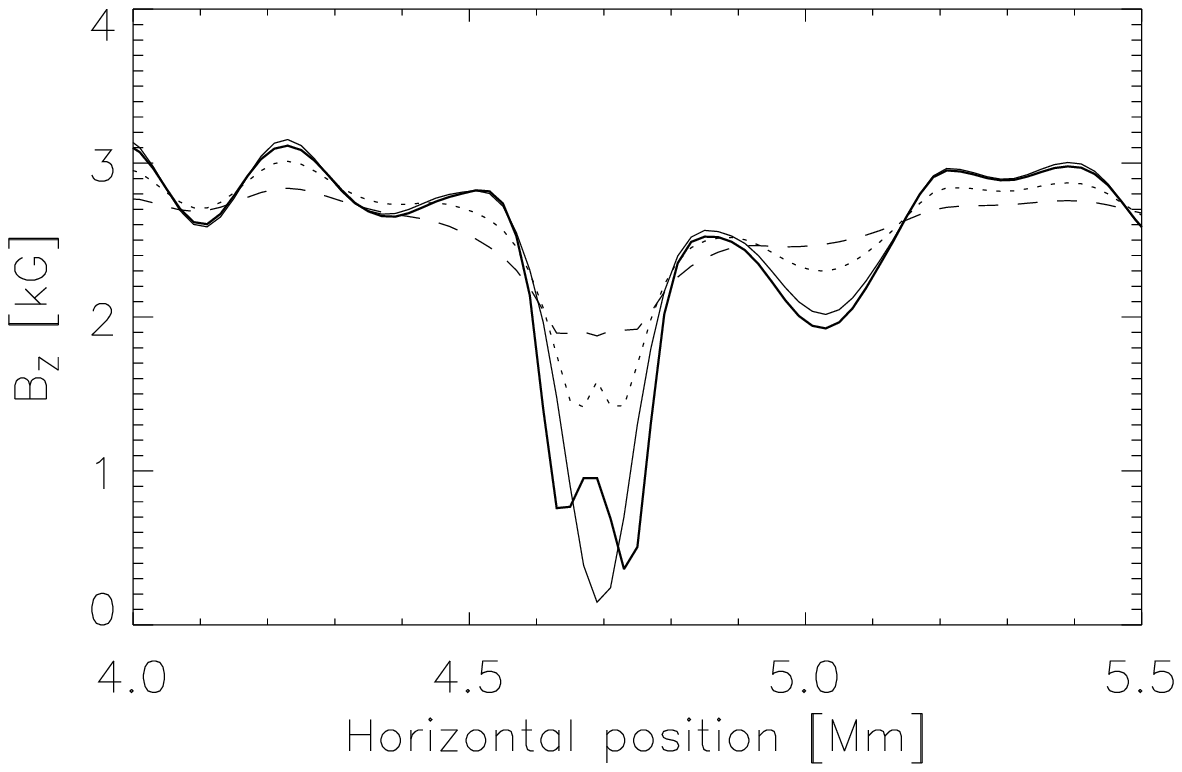}\\
\plotone{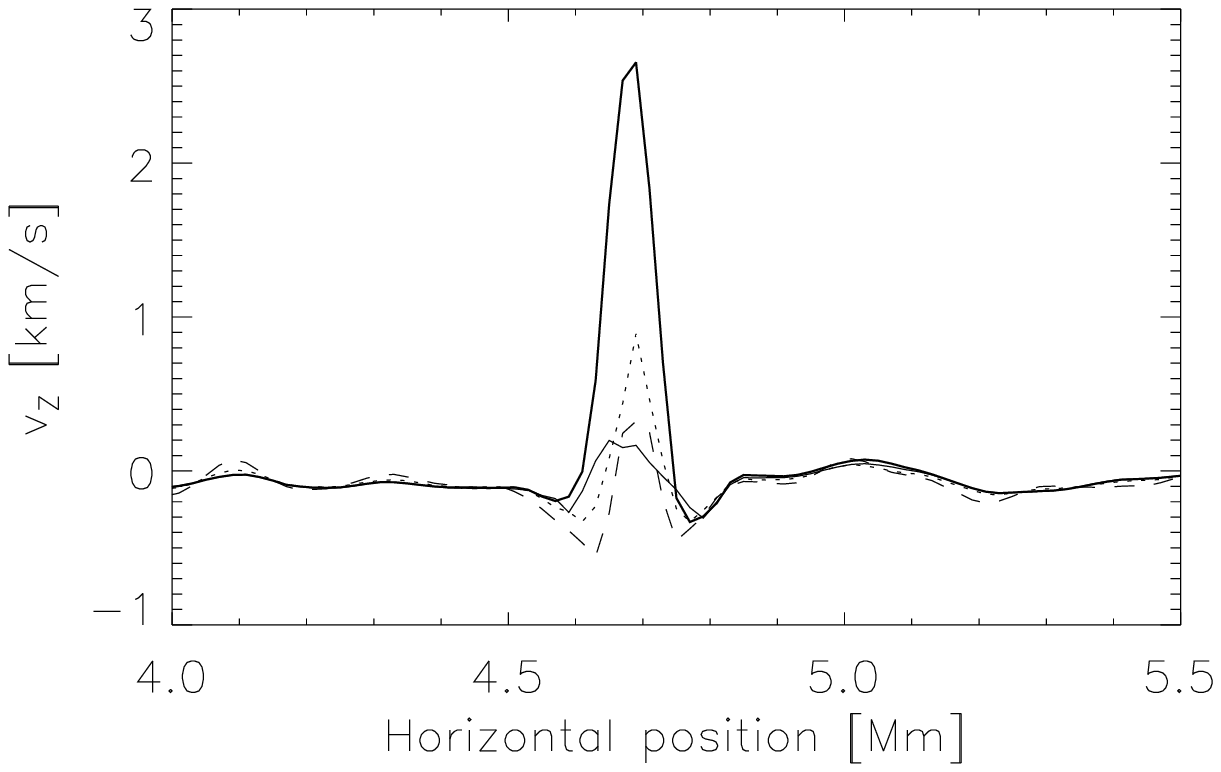}
\caption{Profiles of the vertical magnetic field (upper panel) and
  velocity (lower panel) components along the horizontal direction in
  the cut shown in Fig.~\ref{fig:cut_y}. In both panels, the thick solid
  curve corresponds to constant geometrical height $z=1200\,$km and the
  other three lines indicate the levels of constant (Rosseland) optical
  depth $\tau_{\mathrm R}=1$ (thin solid line), 0.1 (dotted line), and
  0.01 (dashed line).  While the variation between the quantities inside
  and outside the upflow plume (umbral dot) is very large at the same
  geometrical level, it is strongly reduced at equal optical depth.}
\label{fig:cuts_by_vy}
\end{figure}

The profiles of vertical components of velocity and magnetic field along
the horizontal direction of the cut shown in Fig.~\ref{fig:cut_y} are
given in Fig.~\ref{fig:cuts_by_vy}. There is a strong difference between
the values inside and outside the upflow plume at constant {\em
geometrical} level, but the elevation of the lines of constant {\em
optical} depth together with the strong decrease of the differences with
height leads to much smaller variations in the range $\tau_{\mathrm
R}\simeq0.01\dots0.1$, where most photospheric spectral lines are
formed. At $\tau_{\mathrm R}=0.01$ (0.1) we find only a maximum
reduction of the vertical field strength by 20\% (40\%) with respect to
the horizontal mean of $2500\,$G and a maximum upflow velocity of 0.3
(0.9) km$\cdot$s$^{-1}$.  Consequently, spectroscopic observations would
reveal neither the strong field reduction nor the high velocities in the
upflow plumes, even if any instrumental or seeing-related smearing and
straylight effects were absent. This is consistent with existing
observational results \citep[e.g.,][]{Socas-Navarro:etal:2004}.

Fig.~\ref{fig:timeseries} shows the time evolution of a typical upflow
plume developing into an umbral dot when hot plasma reaches the surface
($\tau_{\mathrm R}=1$). The development starts as a feeble upflow below
the surface (left panels), which develops into a strong plume once the
hot gas reaches the surface, cools by radiative losses, and descends in
adjacent narrow downflow channels, thereby forming an overturning cell. The
expanding rising plasma creates a `gap' in the magnetic field which
develops into a cusp-like configuration outlined by the dark lane in the
brightness images.  The jet- or sheet-like upflow above the cusp driven
by the steeper pressure gradient of the piled-up plasma below is clearly
visible in the velocity cuts. Since there is no sustained buoyancy
driving from below to supply the strong upflow in the upper layers
driven by radiative losses, density and pressure decrease and the
magnetic field strength grows in the deeper parts and,
eventually, the plume fades away.

\section{Discussion}

Our simulations show that upflow plumes and umbral dots naturally appear
in strong-field magneto-convection with radiative transfer. The plumes
start off like oscillatory convection columns below the surface, but
turn into narrow overturning cells driven by the strong radiative
cooling around optical depth unity, which leads to descending
low-entropy fluid. Even though the quantity $\zeta=\eta/\kappa$ is much
smaller than unity in the upper third of our computational
box\footnote{Since our numerical scheme includes a spatio-temporarily
varying thermal `hyperdiffusivity' \citep{Voegler:etal:2005}, the value
of $\zeta$ is not readily determined in the lower part of the
computational box, where $\kappa$ is no longer dominated by the
radiative diffusivity. In the simulations discussed here, the effective
value of $\zeta$ is somewhat larger than unity in most of the lower two
thirds of the box.}, this mode dominates because the overturning flow is
largely confined to the gap in the magnetic field created by the flow,
so that no significant cross-field motion is involved. Extended
calculations in 2D have shown that (a) the character of the
magneto-convective pattern and energy transport as described above is
unaffected if we halve the magnetic diffusivity, so that the results can
considered to be robust in that sense, and (b) for a run with a four
times larger value of $\eta$, the magneto-convection is dominated by an
overturning mode with slowly evolving broad cells and cross-field flow,
similar to what is predicted by linear analysis.

Our simulations reproduce the general properties of a sunspot umbra with
umbral dots. The bright features resulting from the upflow plumes can be
identified with (central) umbral dots; their brightness, size, and
lifetime correspond well to the observations
\citep[e.g.,][]{Socas-Navarro:etal:2004}. The large velocities and small
field strengths in the upflow plumes are effectively hidden from
spectroscopic observations by the pile-up of plasma in the upper layers
of the plumes, raising the formation heights of spectral lines to layers
located largely above the plume. A similar underlying geometry of umbral
dots has been suggested by \citet{Degenhardt:Lites:1993a}; their
assumption of a stationary, unidirectional vertical flow is not
supported by our simulation results, however.

The results show that no intrusion of field-free plasma from the
underlying convection zone into a cluster-type shallow sunspot is
required explain the existence of umbral dots. In fact, a simulation
with an about 15\% larger value of the internal energy of the inflows
resulted in flux separation, i.e., the formation of large field-free
patches with granule-like convection surrounded by very strong, largely
homogeneous magnetic field \citep[cf.][]{Weiss:etal:2002}, a
configuration very much different from that of an undisturbed inner
umbra. Therefore, a hypothetical fragmentation of the magnetic field
into a cluster of flux strands has to occur at a level significantly
below our bottom boundary (located about 1.6 Mm below the solar
surface). However, such intrusions on a larger scale could possibly be
relevant for the formation of light bridges as well as for peripheral
umbral dots connected to bright penumbral filaments.

We have also run a simulation with a field of $3500\,$G and the original
value of the internal energy carried by the inflows. In this case,
umbral dots very rarely occur and the average energy output (per unit
area) is reduced to only about 9\% of its value in the quiet Sun. This
simulation could possibly represent the observed dark cores of sunspot
umbrae, which are largely devoid of umbral dots.


\bibliography{ms}

\clearpage

\clearpage

\clearpage

\end{document}